# Evaluating the Long-Term Viability of Eye-Tracking for Continuous Authentication in Virtual Reality

Sai Ganesh Grandhi and Saeed Samet

School of Computer Science, University of Windsor, Windsor, Canada

## Abstract

*Traditional authentication methods, such as passwords and biometrics, verify a user's identity only at the start of a session, leaving systems vulnerable to session hijacking. Continuous authentication, however, ensures ongoing verification by monitoring user behavior. This study investigates the long-term feasibility of eye-tracking as a behavioral biometric for continuous authentication in virtual reality (VR) environments, using data from the GazebaseVR dataset. Our approach evaluates three architectures—Transformer Encoder, DenseNet, and XGBoost—on short- and long-term data to determine their efficacy in user identification tasks. Initial results indicate that both Transformer Encoder and DenseNet models achieve high accuracy rates of up to 97% in short-term settings, effectively capturing unique gaze patterns. However, when tested on data collected 26 months later, model accuracy declines significantly, with rates as low as 1.78% for some tasks. To address this, we propose periodic model updates incorporating recent data, restoring accuracy to over 95%. These findings highlight the adaptability required for gaze-based continuous authentication systems and underscore the need for model retraining to manage evolving user behavior. Our study provides insights into the efficacy and limitations of eye-tracking as a biometric for VR authentication, paving the way for adaptive, secure VR user experiences.*

## 1. Introduction

The most prevalent authentication method today is static authentication, which includes passwords, biometrics, PINs, and more. While techniques like these are challenging to by pass due to their complex identity structure, static authentication has a noteworthy drawback. Once a malicious actor successfully navigates through static authentication, they gain unrestricted access to the system, which continues to recognize them as a legitimate user[1].

In contrast, continuous authentication offers a dynamic alternative that uniquely combines ongoing user identification with a seamless user experience. Traditional authentication methods typically verify the identity only at the beginning of a session, assuming 1 that the user remains unchanged throughout the session. However, continuous authentication monitors user behavior in real time, making it considerably more challenging for unauthorized individuals to maintain access. Any deviations from established behavioral patterns can trigger immediate responses, such as reauthentication or session termination. This proactive approach significantly mitigates the risk of session hijacking, in which an attacker could take control of an already authenticated session[2].

The evolution of continuous authentication has progressed from desktop environments to mobile devices and now extends to virtual reality (VR) headsets. These headsets utilize behavioral biometrics such as eye tracking, hand movements, and head gestures. Recent studies, including the work by Lohr et al. (2022)[3] and their 2024 follow-up[4], have demonstrated the effectiveness of eye tracking as a behavioral biometric, achieving low Equal Error Rates (EERs) in user identification[3].

Although these advances are promising, behavioral biometric patterns can change over time. It happens due to users changing behavioral patterns; as noted in various studies, behavioral patterns are known to evolve[5]. When behavioral patterns evolve and change, authenticating impostors versus legitimate users becomes difficult, and as a result, EER scores increase. Other behavioral biometrics, such as gaze and touch patterns used in mobile devices, also undergo changes that require continuous updates during the registration phase[5].

In this study, we investigate the long-term usability of eye tracking for continuous authentication using the GazebaseVR dataset, which spans 26 months and encompasses three distinct rounds of data collection. We developed user models based on the transformer encoder architecture utilizing eye-tracking information from Round 1 (the first month) and tested the model's ability to predict user data from Round 3 (captured after 26 months). Initial results showed low accuracy scores of approximately 10 percent. However, when incorporating data from all three rounds into the user model, accuracy scores surged beyond 95 percent. This finding suggests that for eye tracking to serve as a reliable behavioral biometric, the user model must be updated with behavioral data over time.

Our research aims to analyze eye tracking as a viable behavioral biometric for VR/AR headsets by examining its long-term usability. In this study we also explore various authentication architectures—including DenseNet, Transformer Encoder, and XGBoost—to ascertain which algorithms yield optimal results for user identification. Our research findings indicate that while all three architectures achieve commendable accuracy rates, XGBoost demonstrates lower performance with accuracies ranging from 85% to 90%. In contrast, both Transformer Encoder and DenseNet achieve accuracies between 90% and 97%.

## 2. LITERATURE REVIEW

### 2.1. Datasets for Eye Tracking in Virtual Reality

The development of reliable and scalable continuous authentication systems based on eye-tracking requires comprehensive datasets. Several datasets have been introduced in recent years, each offering unique attributes relevant to eye movement biometrics.

#### 2.1.1. GazeBaseVR

GazeBaseVR, developed by Lohr et al. (2023), is a large-scale longitudinal dataset capturing eye movements from 407 college-aged participants using a VR-enabled eye tracker at 250 Hz[6]. Collected over 26 months, the dataset includes 5,020 recordings across five tasks (vergence, smooth pursuit, video viewing, reading, and random saccades), allowing for diverse eye movement analysis. Unique to GazeBaseVR, the dataset provides 3D positional data (X, Y, Z) for both eyes and offers rich demographic diversity, contributing significantly to the field of eye movement biometrics. Compared to its predecessor, GazeBase, this dataset adds novel task types, like vergence, and supports binocular tracking, making it ideal for advanced VR-specific EMB studies. Such extensive data supports the development of robust, generalizable machine learning models for eye movement analysis and authentication applications.

### 2.1.2. GazeBase

The GazeBase dataset, presented by Griffith et al. (2021)[7], is a comprehensive longitudinal dataset featuring 12,334 monocular eye-movement recordings from 322 collegeaged participants[7]. Collected across nine rounds over 37 months, the data includes seven eye-tracking tasks, such as fixation, saccades, reading, and free viewing. All recordings were captured using an EyeLink 1000 eye tracker at 1,000 Hz, with calibration performed for each task to ensure accuracy. Due to its scale and repeated measures, GazeBase is well-suited for studies in eye movement biometrics and machine learning applications focused on eye signal analysis. Additionally, classification labels and pupil area data are available for a subset, providing valuable resources for supervised learning in gaze analysis[7].

## 2.2. Eye Tracking in Continuous Authentication

Eye movement biometrics (EMB) have gained significant attention as potential mechanisms for continuous authentication in VR, where eye-tracking sensors can facilitate real-time identity verification. Recent studies have shown that gaze-driven biometrics can yield low equal error rates (EERs), essential for effective authentication.

### 2.2.1. EKYT and DenseNet Implementation

The Eye Know You Too (EKYT)[3], based on DenseNet architecture, is optimized for eye movement-based biometrics. The EKYT network employs eight convolutional layers with densely connected layers to enhance feature extraction, followed by a global average pooling layer and a fully connected layer, which generates a 128-dimensional embedding for each user. This architecture has demonstrated robust performance for gaze-based continuous authentication, and its DenseNet foundation supports the efficient reuse of features across layers, addressing challenges related to feature extraction in eye-tracking data[3].

### 2.2.2. Gaze Base VR and DenseNet Implementation

The GazeBaseVR dataset collects 5020 binocular eye movement recordings from 407 college-aged participants over three rounds, enabling EMB research in VR environments[6]. Raju et al.[4] have implemented the EKYT architecture in the GazeBaseVR data set. This study contrasts the biometric performance of VR-collected data with a high-end 1,000 Hz eye tracker, showing that while VR data are noisier, it remains viable for authentication, achieving an equal error rate (EER) of 1.67% in short-term scenarios. These findings underline the potential of VR-based EMB, suggesting that VR eye-tracking data, despite challenges, may offer a convenient, accurate biometric solution.

## 2.3. Challenges in Eye Tracking for Continuous Authentication

While gaze-based biometrics hold promise, challenges remain, particularly in terms of calibration, signal quality, and user behavior variation over time. For instance, the visual axis, requiring user-specific calibration, can be challenging for continuous authentication. Raju et al. (2024) noted that spatial accuracy directly affects authentication performance, with higher error rates observed when calibration is not consistently maintained. Further, the dynamic nature of user behavior suggests a need for adaptive models that update with user data to maintain high authentication accuracy over time.

## 3. METHODOLOGY

### 3.1. Dataset

The GazebaseVR dataset stands as the most comprehensive publicly accessible dataset focused on eye-tracking data acquired from virtual reality (VR) and augmented reality (AR) headsets. This dataset encompasses eye-tracking data collected from both eyes of participants while immersed in VR, a setup essential for analyzing human gaze behaviors in virtual environments. The study began with 465 individuals, but 58 were later excluded due to various considerations. The data collection spanned three recording rounds over a 26-month period, with each round incorporating two separate recording sessions approximately 30 minutes apart. The eye-tracking data (ET) were recorded using SensoMotoric Instruments' (SMI's) VR device, which samples data from both eyes at a nominal rate of 250 Hz. Such a high sampling frequency enables precise capture of eye movements, making the dataset ideal for analyzing fine-grained eye movement patterns.

#### 3.1.1. Dataset Tasks

To capture a comprehensive set of eye movement patterns, researchers instructed participants to perform five distinct tasks. These tasks were specifically designed to induce various eye movements such as vergence, smooth pursuit, saccades, and fixations, providing a rich basis for eye movement analysis.

Table1:Overview of Eye Movement Tasks [6]

| Task | Features | Description |
|------|----------|-------------|
| Vergence task(VRG) | Convergence and divergence | A black dot appears on a large Square plane and alternates between different depths. |
| Smooth pursuit task (PUR) | Saccades, fixations | A small black sphere moves Smoothly between the left and right edges of the viewing region. |
| Video viewing task (VID) | Multiple features | A video is displayed on a large, Rectangular plane. |
| Reading task(TEX) | Multiple features | Anexcerptofapproximately820 characters from National Geographic is displayed. |
| Randoms accade task (RAN) | Saccades, fixations | A small black sphere jump storan-Dom screen positions. |

#### 3.1.2. Dataset features

The ET API provided by SMI produces 3-dimensional unit vectors representing the gaze direction of each eye and timestamps with nanosecond precision. There are 250 timestamp records for each second (250Hz), which provides a rich analysis of eye movements [6]. The following features are collected for each user. We only utilize features n, clx, cly, clz, crx, cry, crz and an additional feature created by us called user. The first 7 features provide patterns of eye movements such as fixations, saccades, blink and more of users and these are used to train the user model which is described in further sections. The last column user is used for multi-classification, it basically represents which users information and we classify a user.

## 3.2. Pre-Processing of Eye-Tracking Data

The pre-processing of raw eye-tracking data is essential to ensure consistency and quality for subsequent analysis and model training. This involves selecting relevant features, normalizing data temporally and spatially, and structuring it into segments suitable for model input.

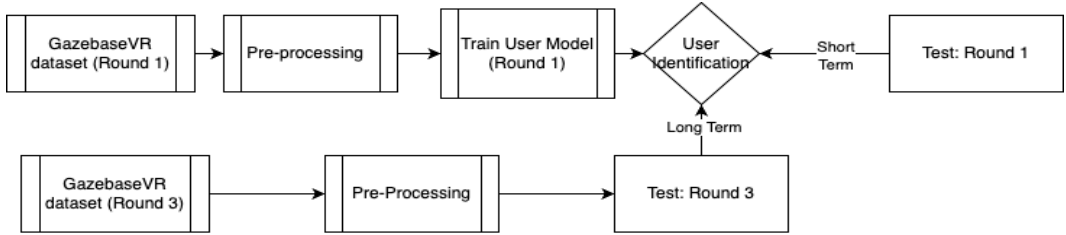

Figure1: Overview of the Methodology

### 3.2.1. Feature Selection

Key features are selected from the raw data to focus on essential gaze patterns. These include the timestamp ('n') and six spatial coordinates: 'clx', 'cly', 'clz' for the left eye, and 'crx', 'cry', 'crz' for the right eye. For temporal normalization, timestamps are converted from milliseconds to seconds, ensuring consistency across devices and enabling standardized time-based pattern analysis[4].

### 3.2.2. Normalization

The normalization process for eye-tracking data in this study involves both temporal and spatial aspects, ensuring consistency and optimal input for model training. Temporal normalization converts timestamps from milliseconds to seconds:

$$t_{normalized} = \frac{t_{normalized}}{1000}$$

Spatial normalization adjusts each coordinate to a range of [-1, 1] using Min-Max normalization:

$$x_{normalized} = 2 \cdot \frac{x - x_{min}}{x_{max} - x_{min}} - 1$$

where x is the original value and xmin and xmax are the minimum and maximum values across the dataset. The range is centered around zero, which can be beneficial for algorithms like Transformers, which those using activation functions. [8].

### 3.2.3. Windowing and Data Structuring

To prepare the continuous eye-tracking data for model input, we employ a segmentation strategy adapted from the DenseNet architecture [3]. The data stream is divided into fixed-size windows, each containing 1250 samples. This window size corresponds to a 5-second interval when sampled at 250 Hz, striking a balance between capturing meaningful temporal patterns and maintaining a manageable input size for machine learning algorithms. The data within each window is then restructured into a 2D array format. In this arrangement, rows represent the spatial coordinates of eye movements, while columns correspond to discrete time points. This organization results in a 3D array structure (windows × coordinates × time points), which is particularly well-suited for sequential data processing in transformer models [9].

## 3.3. Transformer Encoder Architecture

The proposed architecture, known as the Transformer Model, is specifically designed for processing eye movement biometrics (EMB). This model performs a mapping $f : R(C×T) \rightarrow RN$, where C represents the number of input channels, T denotes the length of the input sequence and N corresponds to the number of output classes. The architecture is inspired by the original Transformer model [9] and has been adapted to efficiently handle the time-series data inherent in eye tracking applications.

The Transformer Encoder architecture has not previously been applied to continuous authentication using eye tracking as a behavioral biometric. This study seeks to benchmark its performance against state-of-the-art models like EKYT and DenseNet [3]. Our findings reveal that the Transformer Encoder achieved strong results and, in some instances, surpassed the DenseNet architecture.

The Transformer Model employs a dimensionality of dmodel = 64, utilizes 4 attention heads, and comprises 2 transformer encoder layers. To mitigate over fitting, dropout is systematically applied throughout the architecture. This design ensures that the model remains compact and efficient, making it suitable for deployment in resource-constrained environments such as virtual reality (VR) and augmented reality (AR) devices.

The choice of dmodel = 64 was made to maintain a compact model size while preserving sufficient representational capacity. This dimensionality allows for efficient processing of the eye-tracking time series data, which consists of 7 features (timestamp and 3D coordinates for each eye) sampled at 250 Hz. The model uses 4 attention heads and 2 encoder layers. This configuration was chosen to capture complex temporal dependencies in eye movement patterns while keeping the model lightweight.

### 3.3.1. Training Parameters

The model is trained for 50 epochs. This number was chosen to provide sufficient iterations for the model to learn patterns in the eye-tracking data while balancing computational resources. An initial learning rate of 0.001 is used with the Adam optimizer. This relatively low learning rate was selected to ensure stable training, particularly important for transformer models which can be sensitive to learning rate. A batch size of 32 is employed, striking a balance between computational efficiency and providing sufficient stochastic gradient estimates. These parameter choices reflect a balance between model complexity and computational efficiency, tailored to the specific requirements of continuous authentication in VR environments using eye-tracking data.

## 3.4. Model Training for Eye-Tracking Data

The model training process involves distinct steps for each of the three models: XGBoost, Transformer Encoder, and DenseNet. Each model is tailored to leverage eye-tracking data for user identification, utilizing varying architectures and methodologies to capture unique gaze patterns. The primary motivation for training XGBoost is its lower computational cost compared to neural network-based approaches. This study aims to evaluate whether computationally efficient machine learning algorithms can deliver comparable performance to neural network models.

### 3.4.1. Training XG Boost Model

The XGBoost model, widely recognized for its gradient boosting capabilities, is used here with a flattened 2D input derived from the original 3D eye-tracking data. The data is first reshaped to a 2D format, allowing XGBoost to process it as a feature matrix where each row represents a user sample. XGBoost is highly suitable for tabular data due to its ability to optimize complex feature interactions through gradient boosting [10]. For 7 model evaluation, the dataset is split in an 80:20 ratio, with the training set comprising 80% of the data and the test set comprising 20

The training objective is set to "multi:softmax" for multi-class classification, with a learning rate (η) of 0.3 and maximum depth (max depth) of 6. A DMatrix is created for each dataset split, providing a structured way for XGBoost to handle labeled data. After training, predictions are made on the test set

This approach capitalizes on XGBoost's strength in handling non-linear data interactions, delivering high accuracy on gaze-based classification tasks.

### 3.4.2. Training Transformer Encoder Model

The Transformer Encoder model is designed to leverage the sequential nature of eyetracking data, using attention mechanisms to capture temporal dependencies and spatial relationships in the gaze patterns. In this setup, the pre-processed eye-tracking data is first converted to PyTorch tensors, with user labels encoded for classification. The data is split into training and testing sets, with batches handled by PyTorch's 'DataLoader' to optimize memory and computation.

The model architecture consists of an embedding layer that maps the input to a dmodel dimensional space, followed by a positional encoding layer to account for sequence order. The encoder structure includes multi-head self-attention and feed forward layers, following the formula:

$$\text{Attention}(Q, K, V) = \text{softmax}\left(\frac{QK^T}{\sqrt{d_k}}\right) V$$

where Q, K, and V represent query, key, and value matrices, respectively, and dk is the dimensionality of the keys. The model is trained for 50 epochs with Cross Entropy Loss as the criterion, and parameter updates are managed by the Adam optimizer with a learning rate of 0.001 [9]. After each epoch, average loss is recorded to track model convergence. This model effectively learns sequential dependencies in eye movements, a crucial feature for reliable user identification.

### 3.4.3. Training DenseNet Model

The DenseNet model, adapted from a convolutional neural network (CNN) structure, is tailored for eye-tracking data through dense connections that encourage feature reuse and efficient gradient flow[11]. The model architecture begins with an initial convolutional layer, followed by multiple dense layers, where each layer receives input from all preceding layers within the dense block. This connectivity enhances learning efficiency and mitigates the vanishing gradient problem, a common issue in deep CNNs.

The DenseNet model processes each input sample in a 1D convolutional format, with dense blocks of increasing dilation rates, allowing it to capture spatial dependencies across varying scales. The loss function used is Cross Entropy Loss, and the Adam optimizer updates parameters to minimize classification error. Similar to the Transformer model, DenseNet is trained for 50 epochs, with loss values logged for each epoch.

DenseNet's structure is advantageous for gaze data, as its dense connections capture both fine-grained spatial details and broader contextual information, contributing to high performance in gaze-based user identification tasks.

## 4. RESULTS AND DISCUSSION

### 4.1. Short-Term Model Results

The initial phase of the experiment focused on short-term model performance, where XGBoost, Transformer Encoder, and DenseNet models were trained on Round 1 eyetracking data of 407 users and evaluated on the same round, using an 80:20 train-test split. Table 2 shows the accuracies achieved for different eye movement tasks, demonstrating that both transformer and DenseNet models performed exceptionally well in classifying users based on their gaze patterns.

In this short-term scenario, the models achieved accuracies between 80.25% and 97.77% across various tasks, with few exceptions going below 80%, indicating that Transformer Encoder and DenseNet effectively capture unique gaze characteristics over a short timeframe. However, XGBoost is limited in performance, with 79.31% accuracy combined with all tasks. Although the accuracy of XGBoost does not meet with neural network architecture, it still provides exemplary accuracy. On the other hand, high accuracies with neural networks suggest that gaze patterns contain distinct features that can differentiate users with a high degree of reliability when the data collection and testing occur within a relatively close period. For tasks like Vergence (VRG) and Smooth Pursuit (PUR), which involve precise eye movements, the accuracy was exceptionally high, reflecting the stability of these gaze patterns over a short term.

This finding highlights the feasibility of using gaze-based biometrics for short-term authentication in VR settings, where users' gaze patterns remain stable and predictable. The high short-term accuracy also underscores the potential of Transformer-based architectures to handle sequential eye-tracking data effectively. The window size for all tasks is 5 seconds, and data from all 407 users is used in training and testing.

Table2:Model Accuracy Comparison-Short-term Training

| Task | DenseNet | Transformer | XG Boost | Train Round | Test Round |
|------|----------|-------------|----------|-------------|------------|
| All | 97.09% | 97.20% | 79.31% | Round1 | Round1 |
| PUR | 96.61% | 96.80% | 84.16% | Round1 | Round1 |
| RAN | 95.52% | 95.58% | 80.25% | Round1 | Round1 |
| TEX | 90.22% | 91.00% | 57.48% | Round1 | Round1 |
| VID | 87.22% | 90.50% | 57.66% | Round1 | Round1 |
| VRG | 96.47% | 97.77% | 87.39% | Round1 | Round1 |

## 4.2. Long-Term Model Results

To evaluate the long-term stability of gaze-based biometrics, we assessed the model performance by training on the data of Round 1 and testing the data of Round 3 collected 26 months later. As shown in Table 3, this resulted in a significant drop in accuracy, with scores ranging from 1.78% to 10.28% depending on the task and model. This dramatic decrease in performance suggests that gaze patterns are not static and may evolve over 9 time, potentially influenced by factors such as changes in user behavior, eye health, or VR interaction habits. This marked decrease suggests that gaze patterns undergo considerable changes over time[?], presenting significant challenges for maintaining reliable user differentiation in long-term scenarios.

DenseNet demonstrated the highest overall accuracy at 7.79% when combining all tasks, while the Transformer Encoder showed the poorest performance at 3.01%. XGBoost exhibited mixed results, outperforming other models in some tasks like TEX (12.11%) and VRG (11.46%), but underperforming in others such as PUR (4.89%) and VID (1.78%).

The task-specific variations in accuracy suggest that certain gaze behaviors may be more stable over time, while others are highly variable. For instance, tasks related to text reading (TEX) and vergence (VRG) showed relatively higher accuracies, indicating potentially more consistent gaze patterns for these activities. In contrast, the video-watching task (VID) yielded the lowest accuracies across all models, highlighting the complexity of long-term gaze-based user identification, particularly for dynamic visual stimuli[12].

These findings underscore the need for more robust models and feature extraction techniques that can adapt to temporal changes in gaze patterns. Future research should focus on developing methods that can maintain higher accuracy levels over extended periods, possibly by incorporating adaptive learning mechanisms or by identifying more stable, long-term gaze characteristics[12].

The low long-term accuracy indicates that models trained on older data fail to generalize well when tested on data collected after a long interval. For example, the Random Saccade (RAN) and Video Viewing (VID) tasks, which depend heavily on dynamic gaze shifts, experienced substantial performance degradation. The results highlight a critical limitation in the long-term use of gaze patterns for continuous authentication. Behavioral biometrics, like gaze data, are inherently dynamic[13], and this variability over time implies that models trained on gaze data must be periodically updated. This need for frequent retraining or model adjustment aligns with findings in related studies on continuous authentication, where user behavior tends to evolve, leading to potential identification challenges.

Table3:Model Accuracy Comparison-Long-term Testing

| Task | DenseNet | Transformer | XG Boost | Train Round | Test Round |
|------|----------|-------------|----------|-------------|------------|
| All  | 7.79%    | 3.01%       | 4.85%    | Round1      | Round3     |
| PUR  | 10.28%   | 9.94%       | 4.89%    | Round1      | Round3     |
| RAN  | 5.98%    | 7.67%       | 6.15%    | Round1      | Round3     |
| TEX  | 8.40%    | 3.71%       | 12.11%   | Round1      | Round3     |
| VID  | 4.00%    | 2.45%       | 1.78%    | Round1      | Round3     |
| VRG  | 8.06%    | 7.57%       | 11.46%   | Round1      | Round3     |

## 4.3. Revised Long-term Model Results with Updated Data

To address the observed decline in long-term accuracy, a revised model was trained on a combined dataset of Round 1 and Round 3 data. In this setup, the model was tested 10 on previously unused Round 3 data to assess whether incorporating recent data would enhance performance. As seen in Table 4, this approach resulted in significant improvement, with accuracies reaching up to 98.71%, closely matching the short-term results.

These improved results suggest that by continuously updating the training dataset with recent data, the model can better adapt to evolving gaze patterns. This approach, which mirrors periodic retraining, can help maintain high authentication accuracy even as user behavior changes over time. For continuous authentication systems to remain reliable, incorporating recent data into training may be essential, particularly for biometrics subject to temporal variability, like gaze.

This finding underlines the importance of adaptive modeling in the context of continuous authentication. Behavioral biometrics, unlike static identifiers, require flexible models that can accommodate gradual changes in user behavior. Consequently, for eyetracking authentication systems to be feasible in the long term, regular updates with recent behavioral data are likely required. Future research could explore optimal retraining intervals and data selection strategies to achieve a balance between computational cost and authentication accuracy.

Table4:Model Accuracy Comparison-Long-term Training with Updated Data

| Task | DenseNet | Transformer | XG Boost | Train Round | Test Round |
|------|----------|-------------|----------|-------------|------------|
| All  | 98.71%   | 96.52%      | 93.25%   | Round1+3    | Round3     |
| PUR  | 98.50%   | 97.46%      | 88.50%   | Round1+3    | Round3     |
| RAN  | 97.14%   | 98.32%      | 83.66%   | Round1+3    | Round3     |
| TEX  | 98.75%   | 98.22%      | 55.13%   | Round1+3    | Round3     |
| VID  | 86.22%   | 91.78%      | 55.78%   | Round1+3    | Round3     |
| VRG  | 97.17%   | 94.48%      | 87.55%   | Round1+3    | Round3     |

## 4.4. Ethical Implications of Eye-Tracking for Continuous Authentication

While eye-tracking offers promising advancements in continuous authentication for virtual reality environments, it also raises several ethical concerns that must be carefully addressed:

### 4.4.1. Privacy Concerns

Eye movement patterns can potentially reveal sensitive information about a user's mental or physical state. For instance, certain gaze patterns might indicate cognitive load, emotional states, or even medical conditions such as attention deficit disorders or early signs of neurological diseases[14]. It is crucial to ensure that this data is used solely for authentication purposes and not for unauthorized analysis or profiling.

### 4.4.2. Data Security

The secure storage and transmission of eye-tracking data is paramount. Given the sensitive nature of biometric information, robust encryption and data protection measures must be implemented to prevent unauthorized access or data breaches[15]. Developers of VR systems must adhere to strict data protection standards and regularly audit their security protocols.

### 4.4.3. User Consent and Control

Clear and comprehensive user agreements are essential when implementing eye-tracking authentication. Users must be fully informed about what data is collected, how it is used, and who has access to it. Additionally, users should have the option to opt out of continuous authentication or choose alternative methods, ensuring their autonomy in deciding how their biometric data is used[15].

By addressing these ethical considerations, we can work towards developing eye tracking authentication systems that not only enhance security but also respect user privacy, promote inclusivity, and maintain high ethical standards in the rapidly evolving field of virtual reality technology.

## 5. CONCLUSION

This study investigated the use of eye-tracking data as a behavioral biometric for continuous user authentication in virtual reality (VR) environments, with a focus on both short-term and long-term usability. Using the GazebaseVR dataset, we evaluated the performance of Transformer Encoder and DenseNet models for user identification, achieving promising results in short-term experiments with accuracy levels reaching over 97%. These results indicate that gaze patterns in the short term can serve as a reliable biometric, with both the Transformer and DenseNet architectures proving effective in classifying users based on unique eye movement characteristics.

However, when testing model performance over an extended period, significant accuracy degradation was observed, with accuracy dropping to as low as 1.78% for certain tasks after 26 months. This decline highlights a key limitation of behavioral biometrics such as eye tracking: gaze patterns are subject to temporal changes, likely influenced by behavioral shifts, health factors, or user adaptation to VR environments. The findings underscore that, while effective in the short term, static models fail to generalize well over time, making continuous model updates essential for sustaining high accuracy in real-world applications.

To address this, we explored an adaptive model training approach by incorporating recent data into the training set. This method restored accuracy to near short-term levels, with performance improvements exceeding 98%. Such results suggest that periodic model retraining with recent data is crucial to maintaining the viability of gaze-based continuous authentication systems. Adaptive modeling, where data from subsequent sessions are used to update the user model, can potentially offset the temporal variability in gaze patterns, providing a practical solution for long-term user identification in VR.

In conclusion, this study demonstrates the feasibility of using eye-tracking as a behavioral biometric for continuous authentication in VR settings, while emphasizing the need for adaptive model retraining to account for behavioral drift over time. Future work could focus on determining optimal retraining intervals and exploring additional features such as head or hand movements to enhance model robustness. As VR applications grow in importance, developing reliable, adaptive biometric systems for continuous authentication will be essential for enhancing user security in immersive environments.

## ACKNOWLEDGEMENTS

We extend our thanks to the School of Computer Science at the University of Windsor for providing the resources necessary for conducting this research. Special appreciation goes to Dr. Oleg V. Komogortsev and his team for developing and sharing the GazeBaseVR dataset, which played a crucial role in our analysis. Additionally, we are grateful to our colleagues for their constructive feedback and insightful discussions that enriched our work.